%% file: Haag.tex
\documentclass[12pt]{article}
\usepackage{dina4p,amssymb,amsfonts,epsfig}

\newcommand{\fT}{{\frak T}}
\title{An Evolutionary Picture for Quantum Physics}
\author{Rudolf Haag\\
Waldschmidtstr. 4b\\
D-83727 Schliersee-Neuhaus\\
Germany}
\date{}
\begin{document}
\maketitle

\parbox{85mm}{Dedicated to the memory of Misha Polivanov who strove to conserve
  a cultural heritage in dark times\\}

\begin{abstract}
In the orthodox language of Quantum Mechanics the observer occupies
a central position and the only "real events" are the measuring
results. We argue here that this narrow view is not forced upon us
by the lessons of Quantum Physics. An alternative language, closer
to the intuitive picture of the working physicist in many areas,
is not only possible but warranted. It needs, however, a different
conceptual picture ultimately implying also a different mathematical
structure. Only a rudimentary outline of this picture will be
attempted here. The importance of idealizations, unavoidable in any
scheme, is emphasized. A brief discussion of the EPR-phenomenon
is added.
\end{abstract}

\section{Language and philosophical extrapolations}

Prominent in the vocabulary of Quantum Theory are the words "physics
systems", "state", "observable", "measuring result". The general
theory tells us how these terms are represented in the mathematical
scheme and it tells us the following: {\em If a system S is in a state
s and we measure the observable A then the probability of obtaining the
result a is given by the formula $p={\rm tr} sP_a^{(A)}$.} I shall not
explain the formula since you know it all.

This language has proved to be very efficient in a wide area.
Nevertheless we should not consider it as sacrosanct. There are
limits to its usefulness and every word in the vocabulary is
subject to criticism.

Let us start with the word "observable". It suggests that there is
an observer. Does this have to be human being? Certainly in the
discussions of the early days of Quantum Mechanics no other
interpretation was intended. One of the concerns of Niels Bohr
was epistemology i.e. the question of what we (humans) can know
and how we can communicate. But even if we want to understand the word
observer in a wider sense we must endow him at least with the
faculties of consciousness, intelligence in planning and free will
in execution. So there is the question: does Quantum Physics force
us to abandon the old picture of a real outside world, called
nature, which exists separate from our consciousness? Do the
finding of atomic physics decide in favour of some philosophical
system like positivism or idealism in contrast to realism? I do
not think so. The raw material of physics, which the theory is
supposed to explain, consists of facts which can be documented.
Nobody claims that in the recognition of a dot on a photographic
plate or of the print out of a computer the quantum mechanical
uncertainties play any role. What is often claimed is that
documents are necessarily macroscopic and that amplification to
the macroscopic scale is essential for the creation of a fact.
We shall look at this important point carefully below. It has,
however, no bearing on the question about the role of the mind in
the interpretation. No matter what our ultimate philosophical
beliefs are, physics by its very method proceeds from an {\em "as if"}
realism. Thus one can hardly doubt that facts similar to
measuring results occur in nature irrespective of whether they
arise in a planned experiment and enter the consciousness of an
observer. For instance we believe that cosmic rays passing
through a body of water which happens to be at the boiling point
may produce lines of vapour bubbles. A child passing by may wonder
about this phenomenon but probably not even notice it. Thus we
may assume that a measuring result is an event whose reality status
is no better than that of other events in nature.

Granting this we must ask: what constitutes an event in the above
example? There are bubbles marking approximate points in space-time
and we attribute these to an elementary quantum process such as
\begin{equation}
 \mu +\  {\rm atom}\  \rightarrow \  \mu + \  {\rm ion} \  +{\rm e}+\gamma
\end{equation}
which creates a localized disturbance in the superheated liquid
and this in turn acts as a germ of vaporization. Can we separate
the elementary process (1) as a closed process from the subsequent
macroscopic amplification? What if the temperature of the water was
a few degrees lower and no bubbles were formed? This brings us back
to the question about the role of amplification. Clearly we
have to distinguish between documents and facts. While the former
are needed for the unequivocal establishment and communication
of a result of observation i.e. are indispensable on the
epistemological side we should recognize that physics transcends
epistemology. In physics we try to extrapolate from what we know
or even can know to form a coherent picture of the world using
criteria like reasonableness, simplicity {\dots}. Observations are a tool
and a check, not the ultimate purpose. The assumption of (1) as an
individual fact is an idealization which has to be judged by its
reasonableness.

\section{Individuals and ensembles.}

Niels Bohr is sometimes regarded as a crown witness for positivism:
his emphasis on epistemology seems to provide some justification
for this. But Bohr disclaimed such a label and reportedly felt
unhappy about this misunderstanding of the message of Quantum
Mechanics. Indeed in his writings you find no trace of a doubt about
the real existence of electrons and atoms but only about our
ability of assigning simple attributes to them. One central point
of Bohr's argument is that Quantum Theory introduces a discrete
element into physics which implies not only the stability of atoms
but also the indivisibility of quantum processes whether it be a
quantum jump in an atom or the passage of a particle between source and
detector in the double slit experiment. Any subdivision of such a
process, the attempt to describe it as a continuous development
in space and time, cannot have an objective meaning. The
Schr\"odinger equation does not describe the individual process.
It describes the continuous change of probabilities for possible
facts not the fact itself. Similarly the formulation of quantum
mechanical statements quoted at the beginning, which is essentially
due to von Neumann, refers to the statistical behavior in an
ensemble: the individual fact, called "measuring result", remains
unresolved. This calls attention to the division problem. What
can be singled out as an individual? This question applies in
parallel to matter and to events. To say that matter is composed
of atoms and an atom is composed of electrons, protons and
neutrons is obviously a coarse picture. The Pauli principle implies
that the "constituents" cannot be regarded as individuals and
even where this principle does not enter, e.g. in the case of the
hydrogen atom, the composition picture is only an analogy
providing a useful model for approximations. An individual part
of matter becomes precisely defined only as an asymptotic notion
which can be arbitrarily well approximated by isolation. The
simplest such individuals are particles (including stable, compact
objects like a crystal). Their internal structure is rigidly fixed by
(quantum) laws of nature. Similarly an individual event is an
asymptotic notion. The simplest type of event is a collision
process between particles, well isolated from other matter and
closed by the spatial separation of the reaction products. Its
mental decomposition into subevents or "virtual events" (as in
Quantum Field Theory) is a useful model but no individual
existence of the virtual events can be claimed. This hinging of
basic concepts to asymptotic situations which are only
approximately realizable emphasizes the need for idealizations.
Idealizations cannot be avoided if we want to talk about any
subdivision of the universe
\footnote{John Bell's quest for "beables"
 which can be precisely defined under any circumstances and his
 criticism of Quantum Theory on the grounds that it is not enough
 to achieve agreement with experiments "for all practical
 purposes" (FAPP) disregards the possibility that with increasing
 insistence on precision the subdivision of the universe must
 become necessarily coarser and the description less detailed.} 
though this does not necessarily have to be done in terms of particles
and collisions. Considering for instance the regime of an extended
medium of high density we may define an individual event as a
significant deviation from local equilibrium. In the orthodox
interpretation the idealization begins with the cut between the
"quantum system" and the observing instruments described
classically and continues with Bohr's effort to define a "closed
quantum process" as a complete description of the experimental
arrangement, a task which demands judicious judgment as to what
is relevant and what is not. This is well adapted to laboratory
situations when we have to consider both articles and macroscopic
hardware. But it leaves open the task of translating a
description of the apparatus into the mathematical representatives
of the state prepared and the observable measured and it does not answer
the question of why the interaction between apparatus and quantum
system leads in each individual case to a "measuring result". Again
the occurance of such events has to be postulated. The attempt to
explain this as a consequence of the formalism in the theory of
measurement, for instance by the study of decoherence, goes only
part of the way (see below).

So far our only change from the orthodox view has been to replace
the notion of "measuring result" by the more general notion of
an "event" which is considered as a fact independent of the
presence of an observer. This has, however, important consequences.
An event is irreversible. It is the transition from a possibility
to a fact. We are raised in the belief that the fundamental laws
do not stipulate an arrow of time but are invariant under time
reversal. The explanation of the manifest irreversibility of
processes around us is delegated to statistical mechanics which,
starting from fundamental laws invariant under time reversal,
arrives at irreversible behaviour in the macroscopic
domain
\footnote{The miracle by which this is achieved is the following.
 Going over to a "coarse grained" description one finds that different
 macroscopic states have very different statistical weights. Starting
 with a state of low weight it is therefore overwhelmingly probable
 that later on it develops into a state of higher weight. There
 remains the question of why we only want to draw conclusions from
 a given situation at an earlier time to that at later times and not
 vice versa and why we usually encounter the situation that at the
 early time the state has low statistical weight. In a laboratory
 situation the latter circumstance can be attributed to the
 experimenter starting his investigation. On the larger scale it
 must be blamed ultimately on cosmology telling us that observed
 large deviations from equilibrium did not arise from an earlier
 situation closer to equilibrium as a consequence of a large
 fluctuation but from one of still lower weight.}. 
If we believe that this is the only mechanism by which
irreversibility can arise we must conclude that the elementary
process, even if isolated, cannot be regarded as real but needs the
macroscopic amplification before we can talk about a fact. This
argument has played a role in many discussions of the quantum
mechanical measuring process. Niels Bohr refers to it in rather
careful and slightly enigmatic formulations, for instance:
"Far from implying a special difficulty, the irreversible
amplification effects on which the registration of atomic objects
depends remind us of the {\em essential irreversibility inherent in
the concept of observation}". Now, if we do not want to place the
concept of observation into the center of physics, we must ask
ourselves: what would be the natural picture if we claim that
there are discrete, real events on any scale?

Starting from this idea we come almost unavoidably to an evolutionary
picture of physics. There is an evolving pattern of events with causal
links connecting them. At any stage the "past" consists of the part
which has been realized, the "future" is open and allows possibilities
for new events. Altogether we have a growing graph or, using another
mathematical language, a growing category whose objects are events and
whose (directed) arrows are the causal links. We assume further that
the relation to space-time is provided by the events. Each event marks
roughly a region in space-time, the sharpness of which depends on the
nature of the event. No independent localization properties of links
is assumed. The reason for this may be seen in the case of low density
where the scheme can most easily be compared with the customary
quantum theoretical description. In this case the causal links
correspond to particles, the events to collision processes between
them. To attribute localization to a particle between two processes
would contradict basic experiences in Quantum Mechanics as emphasized
by Bohr's concept of indivisibility and mathematically described by
the spreading of the wave packet for the center of mass motion over a
large volume. Thus, after the source event of emission we have roughly
a spherical wave function. It should not be interpreted as relating to
the probability for the changing position of a point-like particle but
rather to the probability for the space-time location of the collision
center in a subsequent event. Only after the realization of this
target event we may (retrospectively) assign an approximate world line
and momentum to the particle. Let us suppose here that customary
space-time in which patterns of events and links can be embedded has
been independently defined \footnote{In a more ambitious analysis one
  might hope to use the geometry of patterns as a substitute for
  space-time.}.  A pattern of events and links prior to a given time
is a history.

The quantum laws concern two aspects. On the one hand they must
determine the intrinsic structure of links and events (for instance
the internal wave functions or structure functions of particles).
On the other hand they must give probability laws for the formation
of specific patterns, including the positions of collision centers.
No attempt will be made here to formulate these laws. In the low
density example they can be adapted from standard procedure in
quantum theory. Let us sketch a strongly simplified version of this
which shows some essential aspects. To each type of link
$\alpha$
(here a type of particle) we have an associated Hilbert space
${\cal H}_\alpha$
and we may consider all the subsequently mentioned spaces as
subspaces of the Fock space generated from the
${\cal H}_\alpha$
of all types. Consider for simplicity "maximal" events
(corresponding to the strongest possible decisions). They specify
their backward links completely. If the event has two backward links
of types
$\alpha$
and
$\beta$
then it selects a specific product vector
$\varphi_\alpha\otimes\varphi_\beta\in {\cal H}_\alpha\otimes {\cal H}_\beta$
and transforms it to a vector in the tensor product space
${\cal H}_\gamma\otimes {\cal H}_\delta\otimes\cdots$
corresponding to the outgoing channel
\footnote{We made the further
 simplifying assumption that the choice of a specific outgoing
 channel is included in the characteristics of the event.}. 
This vector is, however, not a product vector but a linear combination
of such. Its expansion into a sum of product vectors depends on a
choice of bases in the factor spaces. The selection of a particular
product vector is realized only by the subsequent events since links
become established only after both source and target event are
realized. A space-like surface not passing through any event defines
a "subjective past" consisting of the pattern of all earlier events.
Among these events there are saturated ones for which all forward
links are absorbed by some other event inside this subjective past
and there are others still having free valence links for the
formation of future events. To such a subjective past we associate a
state which summarizes the probability predictions for possible
extensions of the pattern to the future. In our simplified picture
the state depends only on the subpattern of the unsaturated past
events. As the space-like surface is shifted to the future the
associated state changes as new events appear. This change,
analogous to the "reduction of the wave packet", corresponds to the
transition from a possibility to a fact. Let us illustrate this
in the example of the figure in which the wavy line indicates the
chosen space-like surface. We are interested in the extension of the
past history by the pattern of events 4 and 5 and the newly
established links. The temporal order of 1, 2, 3 is irrelevant
but it is assumed that no other events of the past can play a role
for the events linked to 3. Events 1, 2, 3 fix unit vectors
\begin{equation}
 \Phi_1\in {\cal H}_\gamma\otimes {\cal H}_1' \quad ; \quad
 \Phi_2\in {\cal H}_\delta\otimes {\cal H}_2' \quad ; \quad
 \Phi_3\in {\cal H}_\alpha\otimes {\cal H}_\beta
\end{equation} 
Events 4 and 5 are represented by the rank-1 operators (in Dirac
notation)
\begin{equation}
 c|\Phi_4 \rangle \langle \varphi_\alpha\otimes\varphi_\gamma| \quad , \quad
 c'|\Phi_5 \rangle \langle \varphi_\beta\otimes\varphi_\delta|
\end{equation}
where the
$\varphi_\lambda$
are specific unit vectors in the subspaces
${\cal H}_\lambda$, ($\lambda =\alpha,\beta,\gamma,\delta$)
and
$\Phi_4,\Phi_5$
unit vectors in the tensor product spaces of the new outgoing
channels. The constant c,c' together with the selection of the
backward ties i.e. the vectors
$|\cdot\rangle$
determine the probability for a single event. Thus the probability
for event 4 is obtained by applying the first operator of (3) to
$\Phi_1\otimes\Phi_3$.
This yields a vector whose square length gives this probability. To
obtain the joint probability for events 4 and 5 we have to apply both
operators of (3) to $\Phi_1\otimes\Phi_2\otimes\Phi_3$
and square the length of the resulting vector. This joint probability
shows correlations even though these events may lie space-like to each
other. They are due to the fact that the two events have backward
causal links to a common source (event 3). Moreover the vector
$\Phi_3$
determined by event 3, does not specify a product vector
$\varphi_\alpha\otimes\varphi_\beta$
before both events are realized and thus it is not possible to assign
individual "states" to the not yet established links. It is this
feature which distinguishes the joint probability for events from
the case of classical correlations which result if there is an
individual state for each link (possibly unknown) and the
correlations are between these states of links. A prime example is the
EPR-phenomenon (see below).

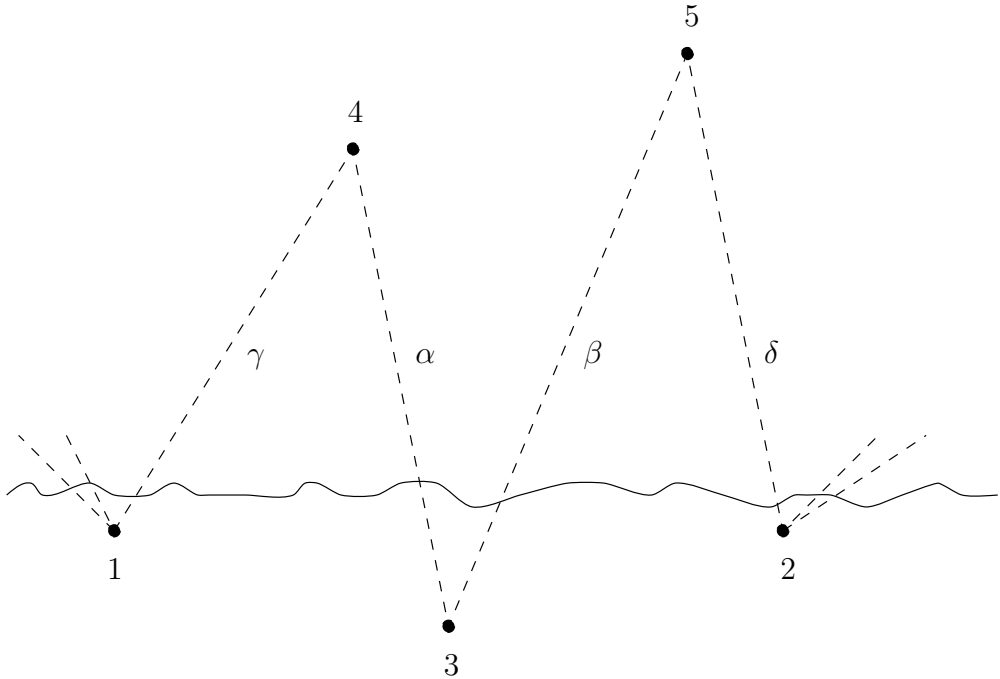
\begin{figure}[hbtp]
\centerline{\input{unnamed.pstex_t}}
\caption{Pattern formation}
\end{figure}

The decision for realization of one possible pattern of events is a
free choice of nature limited only by the probability
assignment.
\footnote{One can speculate whether the decision between
 a large number of alternatives may be decomposed into a sequence
 of binary decisions, each corresponding to one bit of information.
 This would appear indicated in a more fundamental approach.} 
The amount of freedom thus accorded to nature is larger than in the
standard view where the experimenter forces nature to decide only on the answer
to a proposed question. It must be stressed, however, that also
in the standard use of quantum theoretical formalism the element
of free choice by nature cannot be eliminated. It is only pushed
to the rear by focusing on ensembles instead of individual cases.
Thus one may derive from the dynamical law governing the time
development of "states" (representing ensembles) that in the case
of complex systems the density matrix becomes very fast effectively
diagonal in suitably chosen collective coordinates whatever the
initial state may have been. "Effectively" means that in no realistic
experiment the off-diagonal terms will play a role ("decoherence").
One concludes then that this final ensemble {\em may be thought of} as a
mixture of subensembles in each of which the collective coordinates
have specific values. This is perfectly correct as far as
statistical predictions for subsequent measurements are concerned.
It does, however, not explain the fact that in each individual case
nature has decided for one specific set of values (e.g. the position
of a dot on a photographic plate), a decision not controlled by the
experimenter and not described by the time development of the
density matrix. A striking example of the ambiguities involved in
the step from the statistics of an ensemble to conclusions about
individual cases will be discussed below. It is interesting to note
that Dirac advanced the idea of a free choice of nature in this
context in 1927 at the 5th Solvay Congress but was dissuaded by
Bohr who emphasized the decisive role of the observer.

\section{Comparison with standard procedure.}

To compare the degree of compatibility of the scheme with the standard
procedure of Quantum Physics let us first look at a process like (1)
without subsequent amplification effects. The links to the past are
a single
$\mu$
-meson and a single atom far separated from all other matter. In the
conventional treatment we have a Fock space of incoming particles.
The initial state is described as a tensor product of two single
particle wave functions of the respective center of mass motion
(we treat the atom as a single particle). The final state is
described as a vector in Fock space resulting from the application
of the S-matrix to this tensor product. It is a sum of terms
describing the different channels. We write as usual
$S=1+iT$
and, for a particular final channel (suppressing spin indices)
\begin{equation}
 \psi^{out}(p_1'\dots p_n')=\int\tau(p_1'\dots p_n';p_1,p_2)
 \psi_1^{in}(p_1) \psi_2^{in}(p_2) \delta^4(\sum p_k'-\sum p_k) d\mu(p_1)d\mu(p_2)
\end{equation}
with
$d\mu(p_i)=\delta(p_i^2-m_i^2)\Theta(p_i^0)d^4p$.
Using
$$\delta^4(q)=(2\pi)^{-4}\int \exp{ixq}d^4x$$
and noting that
$\exp{ix\sum p_k'}$
represents in any channel just the space-time translation by $x$ in
the Fock space of outgoing particles (similarly
$\exp{ix\sum p_k}$
for the incoming particles) we may write (4) in vector notation as
\begin{equation}
 \Psi^{out}=\int \fT _x\Psi^{in}d^4x,
\end{equation}
regarding this as a mapping in Fock space where
\begin{equation}
 \fT _x=U(x)\fT U(x)^{-1} 
\end{equation}
is the translate by the 4-vector $x$ of an operator
$\fT$
whose matrix elements are the functions
$\tau(p_1'\dots p_k)$.
The latter are smooth functions of the momenta apart from the fact
that they are needed only on the subspace of momentum conservation
and their extension away from this is arbitrary. So we can choose
them to be smooth in all momentum arguments and thereby
$\fT _x$
becomes a quasilocal operator centered 
around $x$. The localization of $\fT _x$ will be poor in the case
of long range forces or "weak processes" like soft
photon emission or interaction with external fields. Let us leave aside
here the problems associated with the existence of massless particles
and focus on hard inelastic events. The characteristics of an event
include the nature of the backward links i.e. the charges, mass
and spin values of the incoming particles and, although they should
not include detailed information about forward links since these are
fixed only in subsequent events, we may include in our case the
choice of a specific final channel and even some rough specification
of the momenta of outgoing particles since this concerns mutually
exclusive possibilities, provided the isolation is adequate.

In the last section we demanded that we should be able to attribute
a rather sharply defined space-time region to the event. This is not
yet provided by the sharpness of localization of
$\fT _x$
(which corresponds roughly to the range of the interaction) but
requires that if we make a cell division in $x$-space, writing
\begin{equation}
 \int \fT _xdx=\sum \fT _k \ ;\quad
 \fT _k=\int \fT _xg_k(x)dx\ ;\quad \sum g_k(x)=1 \ ;\
 \Psi_k=\fT _k\Psi^{in}
\end{equation}
with the function
$g_k$
having support in the cell $k$, then, for appropriately chosen cell
division, the individual terms
$\Psi_k$
may be considered as describing (incoherent) alternatives, one of
which is selected by nature in the individual case. By contrast,
believing in the absolute validity of the quantum theoretical
formalism, one concludes that the phase relation of different
$\Psi_k$
can be put in evidence or, in other words, that the needed size of
the cells depends strongly on far away circumstances surrounding
the process, not only on the event itself (i.e. on the presence of
instruments which are far away at the time of the event). To assess
the significance of this difference we have to study the statistics
of an ensemble of such processes followed by subsequent measurements
on the final state. The relevant test experiment is a very precise
control of the energy-momentum of all initial and final particles.
The assumption of an extension
$a_\nu$
of the event in the
$\nu$
-direction implies a limitation in the control of the momentum
balance
$\Delta P_\nu$
of order
$h/a_\nu$.
This raises the  question of how precisely the relevant part of
past history can be controlled in all samples of the ensemble. Here
the following consideration may be instructive. If the overlap
region of the wave functions of incoming particles were sufficiently
narrow then only a single term
$\Psi_k$
would occur. But this is usually not the case. Consider the
opposite extreme where we take the initial state of the atom in
(1) as an equilibrium state in a large vessel so that its position
is almost unknown. If
$\beta$
is the inverse temperature the state can be described by a density
matrix diagonal in the momentum representation, given by (non
relativistically)
\begin{equation}
 \langle {\bf p'}|\rho |{\bf p}\rangle =\delta^3({\bf p'}-{\bf p})
 \exp{-\beta {\bf p}^2/2m}
\end{equation}
(we have disregarded the normalization). Now we note that precisely
the same density matrix also arises as a mixture of Gaussian wave
packets, minimal at some time t, with width
\begin{equation}
 \lambda=h(\beta/2m)^{1\over 2}
\end{equation}
and distributed with uniform density in space and time. Numerically,
taking for $m$ the proton mass this gives at a temperature of
1 K
a value
$\lambda =2\cdot 10^{-7}$ cm.
Thus it does not make any difference for the statistics of any
subsequent experiment whether we assume that the initial state is
built up from plane waves or from localized packets of size
$\lambda$.
The origin of this ambiguity is, of course, the non uniqueness
of the decomposition of an impure state and we see here that we
cannot confine attention to decompositions into mutually
orthogonal states because we considered mixtures of packets which
are minimal at different times. We are reminded again of the feature
that the study of statistics in an ensemble allows widely different
pictures for the individual case.

Still, there is no known law of nature which would prevent the
control of the momenta of incoming particles and the measurement of
the momenta of outgoing particles with arbitrary precision. Such an
overall high precision experiment would be, in the standard
language, the complementary one to the well known high energy
experiments where we see by inspection in the individual case the
existence of a collision center from which the tracks of particles
emerge. The precision in the definition of this collision center
may not be great but it is much sharper than the controlled
localization of the incoming particles. At present it must be left
to intuition whether one prefers to believe in some fundamental
limitation of the accuracy of the "complementary" experiment or in
the influence of far separated matter on the extension of the
individual collision region.

Let us turn now to patterns of events and links in the low density
situation. A link, corresponding to a particle, is mathematically
described by an irreducible representation of the total symmetry
group which is the direct product of a global gauge group with the
Poincare group. The vectors in this representation space give the
charge quantum numbers and a wave function for the center of mass
motion and spin orientation. The event is described by a {\em reducible}
representation resulting from the tensor product of the irreducible
representations associated with the backward links, followed by
"quasiprojection" by an operator
$\fT _k$.
After the event this representation is decomposed again into a sum
of tensor products of irreducible representations, each term
corresponding to a specific channel of outgoing particles which
furnish possible links to subsequent events. A new event is realized
by the fusion (tensor product) of such links originating from
different past events. We have been careful so far to speak of
representations, not of vectors in the representation spaces.
The reason is that, in contrast to the simplified picture described
in the last section.
$\fT _k$
is not a rank 1 operator and we can only include so much information
about backward links as corresponds to the characteristics we can
attribute to events. These include the approximate momenta determined
retrospectively from the location of the source event but no
assignment of spin orientation for the links is provided. The
$\tau$
-functions in (4) do not factor in the variables of outgoing particles.
This means that we cannot attribute a specific single particle
state to a free valence link and this implies in turn that we cannot
treat the probabilities for the formation of subsequent patterns
as a classical stochastic process. While this complication is not
very relevant for position patterns in the case where the mean free
path is very large compared to the unsharpness of localization of
events so that all momenta can be taken as rather well defined though
unknown, there is no corresponding mechanism providing a specification
of the state of spin orientation of the individual particle. This is
demonstrated by the experiments concerning the EPR-effect for spin.

We consider the decay of a spin zero particle into two spin 1/2 particles
followed by a measurement of the spin orientation of the two
particles with respect to respective directions
${\bf e_1,e_2}$
prescribed by Stern-Gerlach magnets. This may be idealized as the
situation pictured in the figure where events 1 and 2 correspond
to the setting of the Stern-Gerlach magnets, event 3 to the decay
process and events 4 and 5 to the interaction between the decay
particles and the two Stern-Gerlach arrangements each allowing
only a binary decision whose results are denoted by $+$ or $-$. Since
the events 1 and 2 concern the setting of classical apparatus the
links
$\gamma$
and
$\delta$
are already fixed by these events and may be characterized by the
directions
${\bf e_1,e_2}$.
Disregarding the motion in space and focusing only on the spin,
the vector
$\Phi_3$
is the unique singlet state in the Hilbert space of 2-particle spin
states. If 4 is the event with outcome + then
$\varphi_\alpha$
is realized as the single particle state
$\varphi_+({\bf e_1})$
(spin oriented in the
$+{\bf e_1}$
direction). Since the arrangement is such that we are sure that
one of the results $+$ or $-$ must happen the constants c and c' are
equal to 1. The joint probabilities are thus given by the well
known quantum mechanical expressions.

\section{Concluding remarks}

The conceptual structure proposed above incorporates the essential
message of Quantum Physics and does not seem to be at odds with
known experimental findings. The only point of disagreement with
the standard mathematical formalism is the assumed relation
between events and space-time. The clarification of this issue
will demand a considerable amount of theoretical work and
possibly also new experiments. One of the reasons in favor of the
presented picture is precisely this point. It seems ultimately
unsatisfactory to accept space-time as a given arena in which physics
has to play. This feature persists even in General Relativity where
a 4-dimensional space-time continuum is a priori assumed and only
its metric structure depends on the physical situation. In particular,
in the absence of all matter and all events there would still remain
this continuum, void of any significance. This aspect provided
one of the motivations of the author for introducing the notion
of "event" as a basic concept with the ultimate aim of
understanding space-time geometry as the relations between events
\cite{1}. The other motivation was, of course, the desire to
separate the laws of Quantum Physics from the presence of an
observer \cite{2}. In this respect it appears that theorists
discussing quantum processes inside a star or in the early universe
necessarily transcend Bohr's epistemology. Usually then the
orthodox interpretation is silently ignored but there are some
efforts to build a rational bridge from the standard formalism
to such areas of physical theory, most prominently the work by
Gell-Mann and Hartle \cite{3}. It uses the concept of
"consistent histories" introduced by Griffiths \cite{4} and
extended by Omn\`es \cite{5}. For a criticism of this concept see
\cite{6}. Its usefulness is restricted by the fact that consistent
histories embodying some established fasts (measuring results) are
highly non unique. This lead Omn\`es to the distinction between
"reliable properties" and truth.

Still another motivation comes from the following consideration.
The general mathematical structure of standard Quantum Theory is
extremely flexible. Its connection to physical phenomena depends
on our ability to translate the description of circumstances
(e.g. experimental apparatus) to a specification of operators in
Hilbert space. Apart from the case of very simple systems the
success in this endeavor is due to the fact that for most purposes
no precise mathematical specification is needed. Thus, for the
treatment of collision processes in Quantum Field Theory it
suffices to give a division of "all" observables into subsets
which relate to specified space-time regions. However, in addition
to this classification of observables one uses the postulate of
strict relativistic causality. The consequences of this postulate
have been verified by the check of dispersion relations to regions
with an extension far below
$10^{-13}$ cm.
On the other hand it seems highly unlikely that the construction of
an instrument of, say, intrinsic size of
$10^{-15}$ cm
and the control of its placement to such an accuracy could be
possible even in principle i.e. that we may assume the existence of
such {\em observables}. But is it not unlikely that we can attribute to
high energy {\em events} a localization of this order of magnitude
though we have no means of verifying this in the individual dase.
Thus the indirect check by means of dispersion relations could be
explained by the existence of sharply localized events rather than
sharply localized observables.

The realization of a specific result in each individual
measurement has been recognized by many authors as a challenge to
the theory of measurement which cannot be explained using only the
dynamical law of Quantum Theory applied to the interaction of a
quantum system with a macroscopic device but needs an additional
postulate. In the words of Omn\`es this is "a law of nature unlike
any other". In a series of papers Blanchard and Jadczik suggested
a formalism in which irreversibility is introduced in the dynamics
of the coupling of a quantum system with a classical one and thereby
obtained a (phenomenological) description of this aspect of
measurements (see e.g. \cite{7}).

Coming to the evolutionary picture I learned that similar ideas
have been presented by A.N. Whitehead already in 1929 \cite{8}.
His writings have influenced philosophers and theologians but few
if any physicists. In physics D. Finkelstein suggested an approach
to the space-time problem based on similar concepts \cite{9}.
C.F. v. Weizs\"acker tried for many years to draw attention to
the fundamental difference between facts as related to the past
and possibilities as related to the future and argued that for this
reason the statistical statements in physics must always be future
directed \cite{10}.

To conclude let me express my conviction that for a fundamental
physical theory of the future the conceptual structure of
standard Quantum Theory is not adequate. This is no basic
disagreement with the epistemological analysis of Niels Bohr but
the recognition that physical theory always transcends the realm
of experience, introducing concepts which can never be directly
verified by experience though they may possibly be shown to be
incompatible with it.

\end{document}

%% file: unnamed.pstex_t
\begin{picture}(0,0)%
\epsfig{file=unnamed.pstex}%
\end{picture}%
\setlength{\unitlength}{0.00083300in}%
\begingroup\makeatletter\ifx\SetFigFont\undefined
\def\x#1#2#3#4#5#6#7\relax{\def\x{#1#2#3#4#5#6}}%
\expandafter\x\fmtname xxxxxx\relax \def\y{splain}%
\ifx\x\y   
\gdef\SetFigFont#1#2#3{%
  \ifnum #1<17\tiny\else \ifnum #1<20\small\else
  \ifnum #1<24\normalsize\else \ifnum #1<29\large\else
  \ifnum #1<34\Large\else \ifnum #1<41\LARGE\else
     \huge\fi\fi\fi\fi\fi\fi
  \csname #3\endcsname}%
\else
\gdef\SetFigFont#1#2#3{\begingroup
  \count@#1\relax \ifnum 25<\count@\count@25\fi
  \def\x{\endgroup\@setsize\SetFigFont{#2pt}}%
  \expandafter\x
    \csname \romannumeral\the\count@ pt\expandafter\endcsname
    \csname @\romannumeral\the\count@ pt\endcsname
  \csname #3\endcsname}%
\fi
\fi\endgroup
\begin{picture}(6241,4185)(2918,-6088)
\put(5701,-6061){\makebox(0,0)[b]{\smash{\SetFigFont{12}{14.4}{rm}\special{ps: gsave 0 0 0 setrgbcolor}3\special{ps: grestore}}}}
\put(7801,-5461){\makebox(0,0)[b]{\smash{\SetFigFont{12}{14.4}{rm}\special{ps: gsave 0 0 0 setrgbcolor}2\special{ps: grestore}}}}
\put(5101,-2611){\makebox(0,0)[b]{\smash{\SetFigFont{12}{14.4}{rm}\special{ps: gsave 0 0 0 setrgbcolor}4\special{ps: grestore}}}}
\put(7201,-2011){\makebox(0,0)[b]{\smash{\SetFigFont{12}{14.4}{rm}\special{ps: gsave 0 0 0 setrgbcolor}5\special{ps: grestore}}}}
\put(7651,-4111){\makebox(0,0)[lb]{\smash{\SetFigFont{12}{14.4}{rm}$\delta$}}}
\put(6526,-4111){\makebox(0,0)[lb]{\smash{\SetFigFont{12}{14.4}{rm}$\beta$}}}
\put(5476,-4111){\makebox(0,0)[lb]{\smash{\SetFigFont{12}{14.4}{rm}$\alpha$}}}
\put(4426,-4111){\makebox(0,0)[lb]{\smash{\SetFigFont{12}{14.4}{rm}$\gamma$}}}
\put(3601,-5461){\makebox(0,0)[b]{\smash{\SetFigFont{12}{14.4}{rm}\special{ps: gsave 0 0 0 setrgbcolor}1\special{ps: grestore}}}}
\end{picture}